# Observation of Andreev bound state and multiple energy gaps in the non-centrosymmetric superconductor BiPd

Mintu Mondal[a], Bhanu Joshi[a], Sanjeev Kumar[a], Anand Kamlapure[a], Somesh Chandra Ganguli[a], Arumugam Thamizhavel[a], Sudhansu S Mandal[b], Srinivasan Ramakrishnan[a*] and Pratap Raychaudhuri[a†]

[a]*Tata Institute of Fundamental Research, Homi Bhabha Road, Colaba, Mumbai 400005, India.*
[b]*Department of Theoretical Physics, Indian Association for the Cultivation of Science, Jadavpur, Kolkata-700032, India.*

**Abstract:** We report directional point contact Andreev reflection (PCAR) measurements on high-quality single crystals of the non-centrosymmetric superconductor, BiPd. The PCAR spectra measured on different crystallographic faces of the single crystal clearly show the presence of multiple superconducting energy gaps. For point contacts with low resistance, in addition to the superconducting gap feature, a pronounced zero bias conductance peak is observed. These observations provide strong evidence of the presence of unconventional order parameter in this material.

PACS: 74.20.Rp, 74.45.+c, 74.70.Ad, 74.20.Mn

---

[*] ramky@tifr.res.in
[†] pratap@tifr.res.in

The discovery of non-centrosymmetric superconductors[1] (NCS) CePt$_3$Si, namely superconductors where the crystal structure does not have a point of inversion symmetry, have attracted widespread attention. In superconductors where inversion symmetry is present, the superconducting order parameter (OP) is characterized by a distinct parity corresponding to either a spin-singlet or a spin-triplet pairing. However in NCS, the lack of inversion symmetry combined with antisymmetric (Rashba-type) spin orbit coupling (ASOC)[2] can cause an admixture of the spin-singlet and spin-triplet pairing[3]. In the simplest situation of a single band contributing to superconductivity, this mixing is expected to give rise to a two component OP. In a real system, the order parameter would therefore have two or more components, depending on the complexity of the Fermi surface, giving rise to unusual temperature and field dependence of superconducting parameters[4,5,6,7,8,9,10,11].

Despite numerous theoretical predictions, experimental evidence of an unconventional superconducting state in NCS has been very few, possibly due to the small spin-orbit coupling. Thus the vast majority of NCS, (e.g. Re$_3$W, Mg$_{10}$Ir$_{19}$B$_{16}$, Mo$_3$Al$_2$C, Re$_{24}$Nb$_5$) display predominantly conventional s-wave behavior and occasionally multiband superconductivity[12,13,14,15]. In some systems such as CePt$_3$Si [ref. 16] and UIr [ref. 17], the study of parity broken superconductivity is complicated by strong electronic correlations and by the coexistence of magnetism. One notable exception is Li$_2$Pt$_3$B, in which penetration depth[18,19] and nuclear magnetic resonance[20] measurements provide evidence for the existence of nodes in the gap function. However, a direct spectroscopic evidence for the presence of unconventional order parameter has not been reported for any of these materials.

In this letter, we report directional Point Contact Andreev reflection (PCAR) measurements on a BiPd single crystal [ref. 21], which is a recent addition to the family of NCS superconductors. BiPd has a monoclinic crystal structure with lattice constants $a$=5.63 Å, $b$=10.66 Å, $c$=5.68 Å, $\alpha=\gamma=90^0, \beta=101^0$. Recent thermodynamic and transport measurements[21] on high quality BiPd single crystals (residual resistivity, $\rho$~0.3μΩ-cm and residual resistivity ratio ~160) revealed that the specific heat jump at $T_c$ is smaller than expected for a BCS superconductor, suggesting the possibility of multiple superconducting order parameters in this material.

Directional point contact Andreev reflection (PCAR) spectroscopy[22], i.e. where the conductance spectra ($dI/dV$ versus $V$) are recorded by injecting current from a normal metal through a ballistic point contact along different crystallographic directions in the superconductor, is a powerful tool to investigate the gap anisotropy in superconductors[23,24]. In this work, PCAR spectra were recorded on a BiPd single crystal by injecting current either along $b$ (I∥$b$) or perpendicular to $b$ (I⊥$b$). The central observation from these studies is the presence of a pronounced zero-bias conductance peak (ZBCP) in both crystallographic directions, which coexists with more conventional gap-like features. Our results strongly suggest that a spin triplet OP coexists with a spin singlet OP in this material.

High quality BiPd single crystal was grown by modified Bridgeman technique (for details about crystal growth see ref [21]). The directional point contact measurements were done on a piece of single crystal cut into a rectangular parallelepiped shaped of size 1mm×1.5mm×2mm which had large well oriented faces on the (010) and (001) planes. The superconducting transition temperature, $T_c$ ~ 3.62K of the crystal was determined by measuring ac susceptibility at 60 kHz using a two coil mutual inductance technique[25]. From the resistivity and specific heat measurements on a similar crystal, we estimate the electronic mean free path[21], $l$~2.4 μm at low temperatures. The quality of the crystal was also confirmed by observing de Haas-van Alphen (dHvA) oscillation [26]. Before doing point contact measurement, the crystal surface was polished to a mirror finish. To make ballistic point contact, a mechanically cut fine tip made from 0.25mm diameter Ag wire was brought in contact with the (010) (I∥$b$) and (001) (I⊥$b$) crystal faces using a differential screw arrangement in a conventional sample-in-liquid $^3$He cryostat. I-V characteristics of the junction formed between the tip and the sample were measured at different temperature down to T=0.32K using conventional 4-probe technique. The dI/dV vs. V spectra was obtained by numerically differentiating the I-V curves. For all spectra reported here, the contact resistance ($R_c$) in the normal state varied in the range $R_c$~1Ω to 30Ω. The corresponding contact diameter estimated using the formula[27], $d = \left(\frac{4\rho l}{3\pi R_c}\right)^{1/2}$ ~ 100-500Å, was much smaller than $l$. Therefore, all our point contact spectra are taken in ballistic limit. To further understand the nature of superconductivity, we have measured the upper critical field ($H_{c2}$) and its anisotropy along two crystallographic axes (H∥$b$ and H⊥$b$) by measuring ac susceptibility as function of magnetic field at different temperatures.

We first concentrate on the PCAR spectra at the lowest temperature. From large statistics, we observe two kinds of PCAR spectra, corresponding to I∥$b$ and I⊥$b$ respectively[28]. Fig. 1(a) and (b) show representative evolution with contact resistance for (*dI/dV*) versus *V* spectra for I∥*b*. Figure 1(d) and (e) shows similar spectra I⊥b. In both directions, the striking feature is the observation of a pronounced zero bias conductance peak (ZBCP) which coexists with more conventional gap-like features in the low $R_c$ contacts ((Fig. 1(a) and 1(d))). In addition, for I∥*b*, clear coherence peaks associated with superconducting gaps are observed around 0.1 meV and 0.4 meV respectively. For I⊥*b,* the corresponding structures are observed at 0.4 meV and 0.8 meV respectively. Both the ZBCP and gap features disappear at the bulk *$T_c$* confirming their superconducting origin. As the contact resistance increased by gradually withdrawing the tip in both directions the ZBCP slowly vanishes and we recover spectra with only gap-like features. To quantitatively obtain the values of the superconducting energy gaps, we fit the spectra using a two-band Blonder-Thinkham-Klapwijk (BTK) model[29,30] generalized to take into account broadening effects. In this model, the normalized conductance ($G(V)/G_N$, where $G_N=G(V>>\Delta)$) is a weighted sum of the conductance of two transport channels ($G_1(V)$ and $G_2(V)$) arising from the two order parameters: $G(V)/G_N = (1-w)G_1(V)/G_{1N} + wG_2(V)/G_{2N}$. $G_1(V)/G_{1N}$ and $G_2(V)/G_{2N}$ are calculated using the generalized BTK formalism using the relative weight factor of the two factors ($w$) superconducting energy gaps ($\Delta_1$ and $\Delta_2$), the barrier potentials ($Z_1$ and $Z_2$) and the broadening parameters ($\Gamma_1$ and $\Gamma_2$) as fitting parameters. All spectra can be fitted very well with this two-band model if we neglect the large ZBCP that arises for contact with low $R_c$. Analyzing more than 50 spectra along I||b and I⊥b (Fig. 1(c) and 1(f)), we observed that the dominant feature is a gap, $\Delta_1$~0.4±0.1 meV present along both directions. For I||b, in a about 50% of the spectra we can clearly resolve a smaller gap, $\Delta_2$~0.1±0.05 meV with $w$~0.2-0.6. On the other hand in 50% of the spectra along I⊥b, we can clearly resolve a larger gap $\Delta_3$~0.8±0.15 meV with $w$~0.1-0.35. We did not obtain any spectra showing the three gaps simultaneously in the same spectra. The large variation in $w$ and the dispersion in gap values arise from surface roughness which limits our inability to precisely inject current along a desired direction. The temperature variation of the superconducting energy gaps are obtained by analyzing the temperature dependence of two point contacts along the two directions with large $R_c$ (Fig. 2(a) and 2(c)), where the ZBCP is suppressed. The temperature dependence of $\Delta_1$, $\Delta_2$ and $\Delta_3$ obtained from a fit[31] of these spectra with the generalized BTK model are shown in Figures 2(b) and 2(d). It is interesting to note $\Delta_1$ has similar temperature variation for both I||b and I⊥b and closes at $T_c$, suggesting that this gap is associated with the same gap function. The small gap $\Delta_2$ observed for

I∥b on the hand decreases rapidly at low temperatures and forms a tail towards $T_c$ as expected for a multiband superconductor.

We first focus on the origin of the ZBCP in the low resistance spectra. Since ZBCP can arise from several origins it is important to analyze the observed ZBCP in BiPd critically. First, we look for extrinsic origins of the ZBCP that are not associated with genuine spectroscopic features. It has been shown that in the case where the point contact is not purely in the ballistic limit, ZBCP can arise from the current reaching the critical current[32] ($I_c$) of the point contact. However, in our case such a possibility can be trivially ruled out for two reasons. First, as we have shown before our contact is well in the ballistic limit even after considering error associated with our determination of contact diameter from $R_s$. More importantly, the conductance spectra at currents larger than $I_c$ cannot contain any spectroscopic information. In our case however, we observe clear signatures of the superconducting energy gap at bias voltages much larger than voltage range where the ZBCP appears. Other origins of ZBCP include (i) magnetic scattering[33,34] (ii) proximity induced pair tunneling (PIPT)[35] and (iii) Andreev bound state[7,36] (ABS) when the superconductor has an unconventional symmetry. Magnetic scattering and PIPT can be ruled out since in the former ZBCP should split under the application of magnetic field and in the latter the ZBCP should get suppressed at small fields of the order of 0.1T. None of these is observed in our measurements[37]. We therefore conclude the ZBCP-s observed here are manifestations of ABS originating from an unconventional component of the order parameter in this material. Further confirmation of the ABS origin of the ZBCP comes from its evolution with contact size. Since the mean size of the ABS is of the order of the dirty limit coherence length ($\xi_0$), the ZBCP originating from ABS gradually disappears as the contact diameter becomes smaller than $\xi_0$. From the upper critical field ($H_{c2}$) measured with H∥b and H⊥b, we estimate $\xi_0$ to be of the order of 20 nm and 17 nm respectively. In figure 3 we plot the height of the ZBCP (defined as the difference between the experimental zero bias conductance and the zero bias conductance obtained from the generalized two-band BTK fit) as a function of $d$ calculated using the Shavrin formula[27]. We can see that for both directions of injection currents the ZBCP gradually disappears as $d \lesssim \xi_0$. We therefore conclude that the ZBCP in BiPd originates from the ABS resulting from an unconventional OP for which the phase varies on the Fermi surface.

We can now put these observations in proper perspective. For a NCS, ASOC takes the form $\alpha \mathbf{g}(\mathbf{k}) \cdot \boldsymbol{\sigma}$, where $\alpha$ is the spin-orbit coupling constant, $\boldsymbol{\sigma}$ the Pauli matrices and $\mathbf{g}(\mathbf{k})$ is a dimensionless vector related to the spin-orbit coupling so that $\mathbf{g}(\mathbf{k})=-\mathbf{g}(-\mathbf{k})$. The ASOC leads to an energy splitting of the two degenerate spin states resulting in two bands for which the spin eigenstates are either parallel or antiparallel to $\mathbf{g}(\mathbf{k})$. When the ASOC is large, the interband pairing is suppressed and hence the superconducting transition temperature is maximized[6] when the quantization direction of the triplet symmetry becomes parallel to $\mathbf{g}(\mathbf{k})$. Therefore, the superconducting gap will consist of a singlet and a triplet component of the form $\Delta(\mathbf{k}) = (\Delta_s(\mathbf{k})\mathbf{I} + \Delta_t(\mathbf{k})\hat{g}(\mathbf{k}) \cdot \boldsymbol{\sigma})(i\sigma_y)$, where $\mathbf{I}$ is the 2×2 identity matrix, $\hat{g}(\mathbf{k})$ is the unit vector pointing along $\mathbf{g}(\mathbf{k})$ and $\Delta_s(\mathbf{k})$ and $\Delta_t(\mathbf{k})$ are the singlet and triplet amplitude of the gap function respectively. This results in two gap functions $\Delta_{\pm}(\mathbf{k})=\Delta_s(\mathbf{k}) \pm \Delta_t(\mathbf{k})$ where each gap is defined on one of the two bands formed by the degeneracy lifting of ASOC. In general both the singlet and the triplet component of the order-parameter can be anisotropic and can change sign over the Fermi surface. An ABS is formed as helical edge mode[11,38,39,40] for each $\mathbf{k}$ when $|\Delta_t(\mathbf{k})| > |\Delta_s(\mathbf{k})|$. In such a situation, $\Delta_-(\mathbf{k})$ can change sign giving rise to nodes in the superconducting gap for one of the bands. While at present we do not know $\mathbf{g}(\mathbf{k})$ appropriate for the monoclinic structure for BiPd, based on the experimental data we propose the following scenario. Since $\Delta_1 \sim 0.4$ meV is observed for both I∥*b* and I⊥*b* and has similar temperature dependence in both directions, this is likely to originate from a one of the gap functions associated with $\Delta_+$. On the other hand $\Delta_2$ and $\Delta_3$ are likely to be both associated with a strongly anisotropic gap function ($\Delta_-$) for which the observed gap values are different for the two different directions of current injection. While in principle $\Delta_1$, $\Delta_2$ and $\Delta_3$ could also arise from a multiband scenario containing three different bands, this is a very unlikely possibility for the following reasons. First, a simple multiband scenario consisting of multiple s-wave gap functions on different Fermi sheets cannot explain the existence of the pronounced ZBCP that we observe in our data. Secondly, we do not observe $\Delta_2$ and $\Delta_3$ simultaneously in any of our spectra despite the surface roughness that produces a significant scatter in their individual for both directions of injection current. It is therefore unlikely that $\Delta_2$ and $\Delta_3$ arise from two different gap functions on different Fermi sheets.

In summary, we report definite evidence of mixing of spin-triplet and spin-singlet OP in the NCS BiPd. Furthermore, the presence of the pronounced ABS observed from the ZBCP suggests that the pair potential associated with the triplet OP is large enough to produce a sign change in at least one of the gap functions. We believe that this observation is an important step towards realizing Majorana Fermionic modes which are predicted to exist in the vortex core of NCS[8]. Search for these Majorana modes using scanning tunneling spectroscopy measurements is currently underway and will be reported in a separate paper.

We thank Daniel Agterberg for valuable feedback on this manuscript.

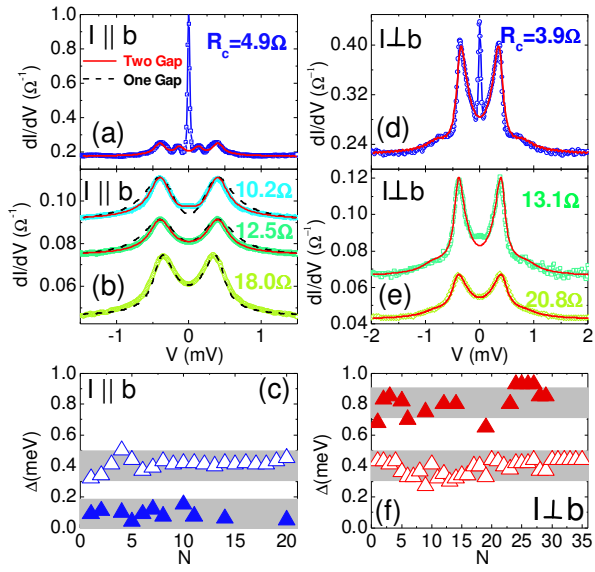

**Figure 1.** (Color online) PCAR spectra for different contact resistance at T~0.35K: (a) and (b) I∥*b* and (d) and (e) I⊥*b*. Solid lines (red) are fits to the modified two gap BTK model. The corresponding fits with a single gap model (black dash lines) are also shown for comparison for some of the spectra. (c), (f) Scatter plot of superconducting energy gap obtained by fitting the modified BTK model to the experimental PCAR spectra for I∥b and I⊥b plotted as a function of the serial number of the spectra.

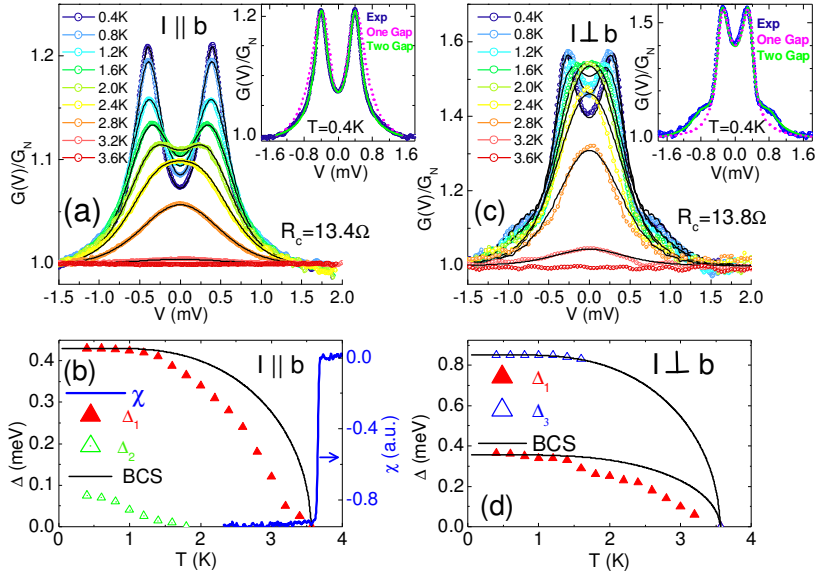

**Figure 2.** (Color online) Temperature dependence of the normalized PCAR spectra of two large resistance contacts for: (a) I∥b and (c) I⊥b. The solid line show the fits to the two-gap BTK model. The inset of fig (a) and (c) show a comparison of single gap and two gap fit to the normalized PCAR spectra at T=0.4K. (b) and (d) Temperature dependence of the superconducting energy gaps extracted from the (a) and (c) respectively. The solid black lines are the expected BCS variation of Δ. In fig (b), the blue solid line shows the ac susceptibility as a function of temperature showing the superconducting transition of BiPd.

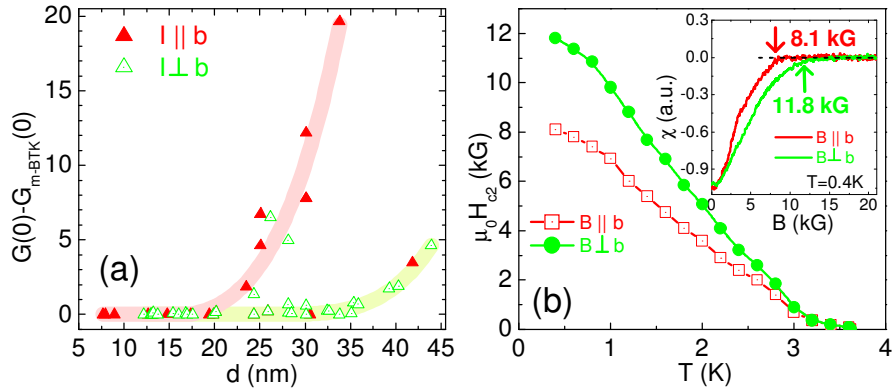

**Figure 3.** (a) Height of the ZBCP as function of contact diameter *d*. The solid triangle (red) corresponds *I*∥*b* and open triangle to *I*⊥*b*. The thick shaded lines are to guide eye. It shows two branches of points associated with ABS in two different directions. (b) $H_{c2}$ as function of temperature (*T*) for H∥*b* and H⊥*b*. The inset shows ac susceptibility as function of magnetic field at T=0.4K. The $H_{c2}$ (shown by arrows) has been extracted from susceptibility data taken as function magnetic field (B) at different temperature.

# Supplementary Material

**Section 1**

As mentioned in the text, the generalized two-band BTK fits contain 7 parameters: The superconducting energy gaps, $\Delta_1$, $\Delta_2$, the barrier parameters $Z_1$, $Z_2$, the broadening parameters $\Gamma_1$, $\Gamma_2$ and the relative weight factor, *w*. In Figs. 2(b) and 2(d) of the main paper we have shown the temperature evolution of the superconducting energy gaps extracted from the spectra obtained with I||*b* and I⊥b. In Figure 2s (below) we show the best fit parameters used for fitting the various spectra at different temperatures. Figure 2s(a) correspond to the parameters used to the fit of the spectra in Fig. 2(a) (I||*b*) and Figure 2s(b) correspond to the fit of the spectra in Fig. 2(c) (I⊥*b*). Figure 2s(c) shows the weight factors corresponding to the spectra in Fig. 2(a) (I||*b*) and Fig. 2(c) (I⊥*b*).

As expected, the values of $Z_1$ and $Z_2$ are nearly temperature independent for both I||*b* and I⊥*b*. The trend in $\Gamma_1$ and $\Gamma_2$ are more complicated. The broadening parameters are formally introduced as an inverse lifetime of the excited quasiparticles. From this perspective one expects this parameter to be small at low temperatures and increase rapidly close to $T_c$ due to recombination of the electron and hole-like Bogoluibons. This is consistent with the temperature variation of $\Gamma_1$ (associated with $\Delta_1 \sim 0.4$ meV) in both Figures 2s(a) and 2s(b). However, $\Gamma_2$ (associated with $\Delta_2$ in Figures 2s(a) and associated with $\Delta_3$ in Figures 2s(b)) in both figures decrease with temperature. To understand this discrepancy we note that phenomenologically the broadening parameter take into account all non-thermal sources of broadening, such as a distribution of gap function resulting from anisotropic gap function (see ref. 23 of the main paper) and instrumental broadening. For a strongly anisotropic gap function, with increase in temperature, intraband scattering can partially smear out the gap anisotropy thereby causing the broadening parameter to decrease. However, the discrepancy could also be an artifact arising from the fact that our fits assume *k*-independent $\Delta_1$ and $\Delta_2$, where the anisotropy of the gap functions are ignored. At the moment we cannot resolve this issue unambiguously.

Figure 2s(c) shows the temperature variation of weight factor *w* for the two set of spectra along I||*b* and I⊥*b*.

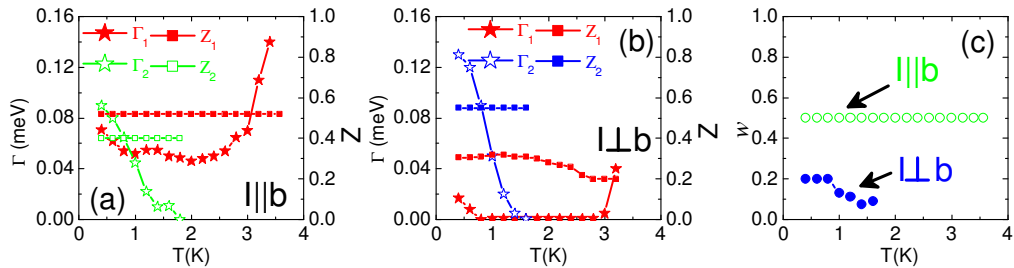

**Figure 2s.** Temperature variation of $Z_1$, $Z_2$, $\Gamma_1$ and $\Gamma_2$ for (a) I ||b and (b) I⊥*b*. (c) temperature variation of *w* for the two sets of spectra along the two directions.

## Section 2

Figure 1s shows the evolution of the Zero Bias Conductance Peak (ZBCP) for two point contacts with I⊥$b$ and I∥$b$ respectively. The ZBCP vanishes in both directions as we approach $T_c$. The magnetic field dependence shows that the ZBCP persists up to moderately high fields of ~0.5 T and does not split with magnetic field. This rules out the possibility of magnetic impurities at the contact or reflectionless tunneling being the origin of this feature.

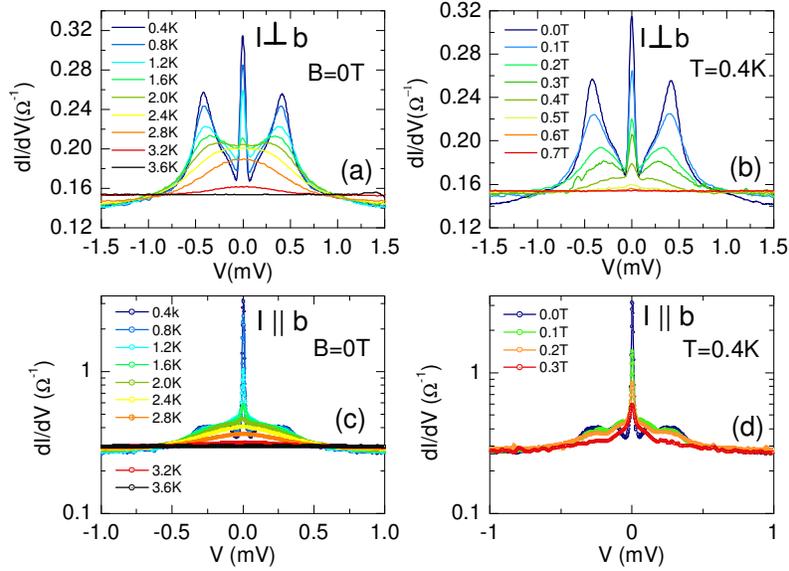

**Figure 1s.** Temperature (a-c) and magnetic field dependence (b-d) of the PCAR spectra for two point contacts with I⊥b and I ∥b showing the ZBCP.